\documentclass[prc,showpacs,showkeys,preprint]{revtex4}
\usepackage{epsfig,graphicx,float,color}
\usepackage{amssymb}
\usepackage[normalem]{ulem}

\begin{document}

\title{Phase diagram of an extended Agassi model}
\author{J.E.~Garc\'{\i}a-Ramos$^{1,2,3}$, J.~Dukelsky$^4$,
  P.~P\'erez-Fern\'andez$^{2,5}$, J.M~Arias$^{2,6}$} 
\affiliation{
   $^1$ Departamento de  Ciencias Integradas y Centro de Estudios
  Avanzados en F\'isica, Matem\'atica y Computaci\'on, Universidad de Huelva,
  21071 Huelva, Spain\\
  $^2$ Instituto Carlos I de F\'{\i}sica Te\'orica y Computacional,
  Universidad de Granada, Fuentenueva s/n, 18071 Granada, Spain\\
  $^3$ Unidad Asociada de la Universidad de Huelva al IEM (CSIC), Madrid,
  Spain\\
  $^4$ Instituto de Estructura de la Materia, CSIC, Serrano 123, 28006
  Madrid, Spain\\  
  $^{5}$ Departamento de  F\'{\i}sica Aplicada III, Escuela T\'ecnica
  Superior de Ingenier\'{\i}a, Universidad de Sevilla, Sevilla,
  Spain\\
  $^{6}$ Departamento de F\'{\i}sica At\'omica, Molecular y Nuclear,
  Facultad de F\'{\i}sica, Universidad de Sevilla, Apartado~1065,
  41080 Sevilla, Spain
}
\begin{abstract}
\begin{description}
\item [Background:] The Agassi model \cite{Agas68} is an extension of
  the Lipkin-Meshkov-Glick model \cite{Lipk65} (LMG) that incorporates
  the pairing interaction. It is a schematic model that describes the
  interplay between particle-hole and pair correlations. It was
  proposed in the 1960's by D. Agassi as a model to simulate the
  properties of the quadrupole plus pairing model. 
\item [Purpose:] The aim of this work  is to extend a previous study
  by Davis and Heiss \cite{Davi86} generalizing the Agassi model and
  analyze in detail the phase diagram of the model as well as the
  different regions with coexistence of several phases.
\item [Method:] We solve the model Hamiltonian through the
  Hartree-Fock-Bogoliubov (HFB) approximation, introducing two
  variational parameters that play the role of order parameters. We
  also  compare the HFB calculations with the exact ones.
\item [Results:] We obtain the phase diagram of the model and classify
  the order of the different quantum phase transitions appearing in
  the diagram. The phase diagram presents broad regions where several
  phases, up to three, coexist. Moreover, there is also a line  and a
  point where four and five phases are degenerated, respectively.  
\item [Conclusions:] The phase diagram of the extended Agassi model
  presents a rich variety of phases. Phase coexistence is present in
  extended areas of the parameter space. The model could be an
  important tool for benchmarking novel many-body approximations. 
\end{description}
\end{abstract}


\maketitle

\section{Introduction}
\label{sec-intro}
Algebraic bosonic and fermionic models with few degrees of freedom, that arose in different areas of physics, served for many years as excellent test beds for many-body approximations appropriate for different areas of interest. They are characterized by a simple Lie group structure \cite{Iach06}  and can be solved either analytically, if a dynamical symmetry is realized, or numerically, for very large system sizes. Let us mention as typical examples, the Jaynes-Cummings \cite{Jayn63} and Dicke \cite{Dick54} models in quantum optics, the LMG model \cite{Lipk65}, the two-level pairing model \cite{Hoga61} and the Elliot SU(3) model \cite{Elli58}, together with the more recent interacting boson model \cite{Iach87} in nuclear physics. The tremendous success of these models let them to permeate  other areas of physics like quantum chemistry, condensed matter and cold atom physics.

More recently, the study of quantum phase transitions (QPTs) and critical points in algebraic models has been an intensive field of research (see, e.g., \cite{Cejn09} and  \cite{Cejn10}). Two of the models that we mentioned above, the LMG model, describing monopole-monopole interactions, and the two-level pairing model were combined in a single model with an SO(5) group algebra by Agassi \cite{Agas68} (see also \cite{Schu69}). The Agassi model has been scarcely used in the literature in spite of its great flexibility and its simplicity to be solved for large systems. Although the random phase approximation (RPA), Hartre-Fock-Bogoliubov (HFB) \cite{Agas68,Schu69, Davi86}  and perturbation theory \cite{Gert83} were applied to this model, modern many-body theories of intensive use in nuclear physics did not profit of the model to asess their applicability and accuracy. As an exception, a recent paper explored the merging of coupled cluster theory (CCT) and symmetry breaking and restoration \cite{Herm17} with the aim posed in future applications to nuclear physics and quantum chemistry.

The Agassi model has a very rich phase diagram explored in Ref.~\cite{Davi86} with a parity broken phase related to the monopole interaction and a superconducting one associated to the pairing interaction. We here extend the Agassi model adding a more general monopole interaction that gives rise to a more complex phase diagram and to several QPTs of different character. We study the model within the mean-field HFB theory and compare with exact diagonalizations in large systems. We also explore the behavior of the appropriate order parameters and derive several critical exponents.

The paper is organized as follows: in Section \ref{sec-model} we
present the algebraic structure of the Agassi model and introduce its
extension with a new term in the Hamiltonian, in Section
\ref{sec-hfb}, the HFB approach is applied to obtain the mean field
energy surfaces and to analyze the stability of the two families of energy
surfaces of the model, in Section \ref{sec-phase-dia}, the
structure of the phase diagram and the nature of the QPTs are
established, and, finally, in Section \ref{sec-comparison}, the
summary and conclusions of this work are presented. 

\section{The extended Agassi model}
\label{sec-model}
The model space of the Agassi model consists in two levels, each of
them with a degeneracy $\Omega$, being $\Omega$ an even number. The
single particle states will be labelled accordingly to the level,
$\sigma=1$ for the upper level and $\sigma=-1$ for the lower one, and
to a magnetic quantum number $m=\pm 1, \pm 2, ..., \pm \Omega/2$,
which labels the states within a given level.  Therefore, the model
space can be physically interpreted as two sub-shells that are part of
a major shell. Hence, $\Omega=2j$, with $j$ an integer number.
Moreover, $\sigma$ can be considered as the parity of the level,
positive for $\sigma=+1$ and negative for $\sigma=-1$. 

The Hamiltonian of the extended Agassi model can be written as,
\begin{equation}
H=\varepsilon J^{0}-g\sum_{\sigma\sigma^{\prime}}A_{\sigma}^{\dagger}
A_{\sigma^{\prime}}-\frac{V}{2}\left[  \left(  J^{+}\right)  ^{2}+\left(
J^{-}\right)  ^{2}\right]  -2hA_{0}^{\dagger}A_{0}.
\label{eq_h_agassi}
\end{equation}
Please, note that the original Agassi model does not contain the last term in (\ref{eq_h_agassi}), $-2hA_{0}^{\dagger}A_{0}$. As we will see, this term introduces new physical effects with respect to the original formulation of the model.

In this work we will redefine the Hamiltonian parameters for convenience, introducing the new parameters $\chi$, $\Sigma$, and $\Lambda$ (see \cite{Davi86}) which are rescaled accordingly to the size of the shell
\begin{equation}
V=\frac{\varepsilon\chi}{2j-1},\quad g=\frac{\varepsilon\Sigma}{2j-1},\text{
\ }h=\frac{\varepsilon\Lambda}{2j-1}.%
\label{eq_par}\end{equation}
We assume the above three parameters as positive because otherwise will lead to unphysical situations. Thus, our extended Agassi Hamiltonian reads,
\begin{equation}
H=\varepsilon \left[ J^{0}- \frac{\Sigma}{2j-1}~\sum_{\sigma\sigma^{\prime}}A_{\sigma}^{\dagger}
A_{\sigma^{\prime}}-\frac{\chi}{2 (2j-1)}~\left[  \left(  J^{+}\right)  ^{2}+\left(
J^{-}\right)  ^{2}\right]  -2 \frac{\Lambda}{2j-1}~A_{0}^{\dagger}A_{0}\right].
\label{eq_h_agassi-si}
\end{equation}
The operators appearing in the Hamiltonian (\ref{eq_h_agassi-si}) are defined as,
\begin{equation}
J^{+}=\sum_{m=-j}^{j}c_{1,m}^{\dagger}c_{-1,m}=\left(  J^{-}\right)  ^{\dagger
},\quad J^{0}=\frac{1}{2}\sum_{m=-j}^{j}\left(c_{1,m}^{\dagger}c_{1,m}%
-c_{-1,m}^{\dagger}c_{-1,m}\right)
\end{equation}

%
\begin{equation}
  A_{1}^{\dagger}=\sum_{m=1}^{j}c_{1,m}^{\dagger}c_{1,-m}^{\dagger},\,
  A_{-1}^{\dagger}=\sum_{m=1}^{j}c_{-1,m}^{\dagger}c_{-1,-m}^{\dagger},\,
  A_{0}^{\dagger}=\sum_{m=1}^{j}\left(  c_{-1,m}^{\dagger}c_{1,-m}^{\dagger
                       }-c_{-1-m}^{\dagger}c_{1,m}^{\dagger}\right)
\label{eq-As}
\end{equation}

\begin{equation}
  A_{1}=\sum_{m=1}^{j}c_{1,-m}c_{1,m},\,
  A_{-1}=\sum_{m=1}^{j}c_{-1,-m}c_{-1, m},\,
  A_{0}=\sum_{m=1}^{j}\left(c_{1,-m}c_{-1, m}-c_{1,m}c_{-1,-m}\right),
\label{eq-As-herm}
\end{equation}

\begin{equation}
N_{\sigma}=\sum_{m=-j}^{j}c_{\sigma, m}^{\dagger}c_{\sigma, m},\qquad
N=N_{1}+N_{-1}.%
\end{equation}
Where $c^\dagger_{\sigma, m}$, $c_{\sigma, m}$ are fermion operators
that create and annihilate a fermion respectively in the
single-particle state ${|\sigma, m\rangle}$. There are $10$
independent generators, $3$ $J$'s, $6$ $A$'s and the particle number,
$N$. Note that $N_{1}$ and $N_{-1}$ are linear combination of $J^{0}$ and $N$. These operators are the generators of the O(5) algebra. 


Therefore, the Hamiltonian (\ref{eq_h_agassi-si}) can be diagonalized
with a O(5) basis \cite{Agas68,Hirs97}. On the other hand, since the Hamiltonian (\ref{eq_h_agassi-si}) commutes with the parity operator $e^{-\imath \pi J^{0}}$, the eigenstates of the system will have either positive or negative parity.

\section{The Hartree-Fock-Bogoliuvob approach}
\label{sec-hfb}
In this section we perform the mean-field energy surface study of the
extended Agassi model. To this end, and closely following
\cite{Davi86}, we will use a  Hatree-Fock transformation followed by a
Bogoliubov one. This approach is well suited for this model in which
the Hamiltonian contains both pairing and monopole
interactions. As shown in \cite{Davi86}, non trivial
BCS and Hartree-Fock broken symmetry solutions are obtained. The Hartree-Fock
transformation can be written as, 
\begin{equation}
a_{\eta,m}^{\dagger}=\sum_{\sigma}D_{\eta\sigma}c_{\sigma,m}^{\dagger}%
\end{equation}
and the Bogoliubov one as
\begin{eqnarray}
\nonumber\alpha_{\eta,m}^{\dagger}&=&u_{\eta}a_{\eta,m}^{\dagger}-sig\left(  m\right)
v_{\eta}a_{\eta,-m},\\
\alpha_{\eta,-m}^{\dagger}&=&u_{\eta}a_{\eta,-m}^{\dagger}+sig\left(  m\right)
v_{\eta}a_{\eta,m},
\end{eqnarray}
where $sig(m)$ stands for the sign of $m$, $sig(m)=+1$ for $m>0$ and $sig(m)=-1$ for $m<0$.

Any calculation in the Agassi model requires to fix the system size, $j$, and the number of interacting fermions. For simplicity we will fix the ratio between the number of fermions and the system size. In the following, we will consider that the number of fermions is $2j$, i.e., the system is half filled and, therefore, there is a number of $j$ fermion pairs. Under this assumption, the following conditions are fulfilled,
\begin{equation}
  u_{-1}^{2}=v_{1}^{2}~~\text{, \ }~~u_{1}^{2}=v_{-1}^{2} ~~ \text{, \ } ~~ v_{\eta}^{2}+u_{\eta}^{2}=1.
  \label{eq_phase}
\end{equation} Therefore, the normal density matrix \cite{Davi86} can be written as,
\begin{equation}
\rho_{\sigma m,\sigma^{\prime}m^{\prime}}=\left\langle c_{\sigma,m}^{\dagger}c_{\sigma^{\prime},m^{\prime}}\right\rangle =\sum_{\eta\eta^{\prime}} D_{\sigma,\eta}D_{\sigma^{\prime},\eta^{\prime}}\left\langle a_{\eta,m}^{\dagger}a_{\eta^{\prime},m^{\prime}}\right\rangle=\delta_{m,m^{\prime}} \rho_{\sigma,\sigma'},
\end{equation}
where
$\rho_{\sigma,\sigma'}=\sum_{\eta}D_{\sigma,\eta}D_{\sigma^{\prime},\eta}v_{\eta}^{2}$. On the other hand, the abnormal density matrix \cite{Davi86}  is
\begin{equation}
\kappa_{\sigma m, \sigma^{\prime} m'}=\left\langle c_{\sigma,m}^{\dagger}c_{\sigma^{\prime} m'}^{\dagger}\right\rangle=\sum_{\eta\eta'}D_{\sigma,\eta} D_{\sigma^{\prime},\eta'} \left\langle a_{\eta,m}^{\dagger}a_{\eta', m'}^{\dagger}\right\rangle = sig\left(m\right) \delta_{m ,-m'}\kappa_{\sigma,\sigma'},
\end{equation}
where
$\kappa_{\sigma,\sigma'}=\sum_{\eta}D_{\sigma,\eta}D_{\sigma',\eta}u_{\eta}v_{\eta}$.

It is possible to write Hartree-Fock and Bogoliubov transformations in terms of only two variational parameters, $\varphi$ and $\beta$, as written below,
\begin{equation}
  D_{1,1}=D_{-1,-1}=\cos\frac{\varphi}{2}\text{ , \ }D_{-1,1}=-D_{1,-1} =\sin\frac{\varphi}{2}
  \label{lasD}
\end{equation} %
and
\begin{equation}
  v_{1}=\sin\frac{\beta}{2}\text{, \ }v_{-1}=\cos\frac{\beta}{2},
  \label{lasv}
\end{equation}
Therefore, the normal density matrix results,
\begin{eqnarray}
\nonumber
  \rho_{1,1}&=&\cos^{2}\frac{\varphi}{2}\sin^{2}\frac{\beta}{2}+
               \sin^{2}\frac{\varphi}{2}\cos^{2}\frac{\beta}{2}=
               \frac{1}{2}\left(1-\cos\varphi\cos\beta\right),\\
  \nonumber
  \rho_{-1,-1}&=&\sin^{2}\frac{\varphi}{2}\sin^{2}\frac{\beta}{2}+
                 \cos^{2}\frac{\varphi}{2}\cos^{2}\frac{\beta}{2}=
                 \frac{1}{2}\left(1+\cos\varphi\cos\beta\right),\\
  \rho_{1,-1}&=&\rho_{-1,1}=\cos\frac{\varphi}{2}\sin\frac{\varphi}{2}\sin^{2}\frac{\beta}{2}
                -\cos\frac{\varphi}{2}\sin\frac{\varphi}{2}\cos^{2}\frac{\beta}{2}
                =-\frac{1}{2}\sin\varphi\cos\beta. \label{lasrho}
\end{eqnarray}
One should note that with the parametrization (\ref{lasD}) and (\ref{lasv}) and conditions (\ref{eq_phase}), two independent phase selections for $u_1$ and $u_{-1}$ are possible: i) $u_{-1}=v_1=\sin\frac{\beta}{2}$ and $u_{1}=v_{-1}=\cos\frac{\beta}{2}$, and ii)  $u_{-1}=v_1=\sin\frac{\beta}{2}$ and $u_{1}=-v_{-1}=-\cos\frac{\beta}{2}$. While the normal density (\ref{lasrho}) does not depend on the phase selection because the coefficients appear squared,  the abnormal density matrix does depend on the phase selected. In particular, using the positive sign for both parameters $u_1$ and $u_{-1}$, case i), one gets
\begin{equation}
  \kappa_{\sigma,\sigma'}=\delta_{\sigma\sigma'}\frac{1}{2} \sin\beta,
\end{equation}
while for the alternative phase selection, ii) above, the abnormal density matrix is,
\begin{eqnarray}
\nonumber
  \kappa_{1,1}&=&\cos^{2}\frac{\varphi}{2}\sin\frac{\beta}{2}\cos\frac{\beta}{2}-
               \sin^{2}\frac{\varphi}{2}\sin\frac{\beta}{2}\cos\frac{\beta}{2}=
               \frac{1}{2}\cos\varphi\cos\beta,\\
  \nonumber
  \kappa_{-1,-1}&=&\sin^{2}\frac{\varphi}{2}\sin\frac{\beta}{2}\cos\frac{\beta}{2}-
               \cos^{2}\frac{\varphi}{2}\sin\frac{\beta}{2}\cos\frac{\beta}{2}=
               -\frac{1}{2}\cos\varphi\cos\beta,\\
  \kappa_{1,-1}&=&\kappa_{-1,1}=\cos\frac{\varphi}{2}\sin\frac{\varphi}{2}\sin\frac{\beta}{2}\cos\frac{\beta}{2}
                +\cos\frac{\varphi}{2}\sin\frac{\varphi}{2}\sin\frac{\beta}{2}\cos\frac{\beta}{2}
                =\frac{1}{2}\sin\varphi\sin\beta.
\end{eqnarray}
Depending on the phase selection, different energy surfaces (A and B) are obtained. This is summarized in Table \ref{phases}.
\begin{table}[h!]
  \begin{center}
    \caption{Bogoliubov phase selection.}
    \label{phases}
    \begin{tabular}{|l|c|}
      \hline
      \hline
      \textbf{Phase selection} & \textbf{Surface} \\
      \hline
      $u_{-1}=v_1=\sin\frac{\beta}{2}$ & \\
      $u_{1}=v_{-1}=\cos\frac{\beta}{2}$& A \\
      \hline
      $u_{-1}=v_1=\sin\frac{\beta}{2}$ & \\
      $u_{1}=-v_{-1}=-\cos\frac{\beta}{2}$& B \\
      \hline
      \hline
    \end{tabular}
  \end{center}
\end{table}
Once the Hartee-Fock-Bogoliubov (HFB) state is defined, with a given phase selection, as a function of the variational parameters, $\varphi$ and $\beta$, the energy surface is obtained as the expectation value of the Hamiltonian (\ref{eq_h_agassi-si})
\begin{equation}
  E(\varphi, \beta) = \frac{\langle HFB(\varphi, \beta)| H  | HFB(\varphi, \beta) \rangle}{\langle HFB(\varphi, \beta)|HFB(\varphi, \beta) \rangle}.
\end{equation}
The surface extrema are studied by minimizing $E(\varphi, \beta)$ with
respect to the variational parameters and, then, analyzing their
stability through the eigenvalues of the Hessian matrix. This is done
in the following two subsections for the two possible phase selections
given in Table \ref{phases}.  

\subsection{Energy surface A}
This energy surface is obtained with the selection of phases as stated
in the first row of Table \ref{phases}.  It can be
written as,
\begin{equation}
E_{A}=-\varepsilon j\cos\varphi\cos\beta-gj^{2}\sin^{2}\beta-Vj^{2}\sin
^{2}\varphi\cos^{2}\beta.
\label{eq-energ1}
\end{equation}

In order to present the results, it is convenient to rescale the energy functional with the size $j$ of the system. Then, the energy functional reads, 
\begin{equation}
\frac{E_{A}}{j\varepsilon}=-\cos\varphi\cos\beta-\frac{\Sigma}{2}\sin^{2}%
\beta-\frac{\chi}{2}\sin^{2}\varphi\cos^{2}\beta,
\label{eq-energ1b}
\end{equation}
being $\varepsilon$ an overall constant of energy (note that the term $-1$ in the denominator of Eqs. (\ref{eq_par}) is not taken into account because a large value of $j$ is assumed). Although the order parameters are $\varphi$ and $\beta$, it is convenient to define combinations of them which are easier to be calculated with a diagonalization. These effective order parameters are
\begin{equation}
\frac{\left\langle J^{+}\right\rangle_A }{j}=\frac{\left\langle J^{-}%
\right\rangle_A }{j}=\sin\varphi\cos\beta,
\label{orpar_e1a}
\end{equation}
\begin{equation}
\frac{\left\langle A_{1}^{+}\right\rangle _{A}}{j}=\frac{\left\langle
A_{-1}^{+}\right\rangle _{A}}{j}=\frac{1}{\sqrt{2}}\sin\beta ~~\text{, \ }~~%
\frac{\left\langle A_{0}^{+}\right\rangle _{A}}{j}=0,
\label{orpar_e1b}
\end{equation}
where the  subindex $A$ refers to the energy  $E_{A}$.

To study the extrema of (\ref{eq-energ1b}), first we impose the
derivatives of the energy surface to be equal to zero 
\begin{eqnarray}
\frac{\partial E_{A}}{j\varepsilon\partial\beta}&=&\sin\beta\left(  \cos
\varphi-\Sigma\cos\beta+\chi\sin^{2}\varphi\cos\beta\right) = 0, \nonumber \\
\frac{\partial E_{A}}{j\varepsilon\partial\varphi}&=&\sin\varphi\cos\beta\left(
1-\chi\cos\varphi\cos\beta\right)=0 .
\label{extrema}
\end{eqnarray}

Later, to determine the nature of the extrema, i.e., minima, maxima or saddle points, we calculate the Hessian matrix (vertical and horizontal lines are included for clarity),

\begin{equation}
  \left(
  \begin{array}{c|c}
     \label{hessian} 
{\cal H}_{\varphi,\varphi} & {\cal H}_{\varphi,\beta} \\
\hline
{\cal H}_{\beta,\varphi} & {\cal H}_{\beta,\beta}
\end{array}
\right) =
\left(
\begin{array}{c|c}
\cos (\beta ) \cos (\varphi ) - & - \sin (\beta ) \sin (\varphi ) + \\
 \chi  \cos ^2(\beta ) \cos(2 \varphi ) & \frac{\chi}{2} \sin (2\beta ) \sin (2 \varphi) \\
\hline
- \sin (\beta ) \sin (\varphi ) +  & \cos (\beta ) \cos (\varphi ) - \Sigma \cos (2 \beta) + \\
     \frac{\chi}{2} \sin (2\beta ) \sin (2 \varphi) &  \chi \cos (2 \beta) \sin ^2(\varphi)
\end{array}
\right) \nonumber
\end{equation}

The solution of the equations (\ref{extrema}) assuming that $\Sigma\neq\chi$ leads to four cases plus a particular case in which $\Sigma =\chi$. These solutions are,
\begin{enumerate}
\item[I-A)]  $\varphi=\beta=0$ ($E_A/(j\varepsilon)=-1)$. Regardless the values of $\Sigma$ and
  $\chi$.

  The Hessian matrix is,
  \begin{equation}
  \left(
  \begin{array}{c|c}
    1 - \chi & 0 \\
    \hline
    0 & 1 - \Sigma
  \end{array}
  \right)
  \end{equation}
 with eigenvalues: $1-\chi$ and $1-\Sigma$. Therefore,
  \begin{itemize}
  \item $\chi<1$ and $\Sigma<1$: it generates a minimum.
  \item $\chi>1$ and $\Sigma>1$: it generates a maximum.
  \item $\chi>1$ and $\Sigma<1$ or $\chi<1$ and $\Sigma>1$: it generates a saddle point.
  \end{itemize}
  Both order parameters are equal to 0. Consequently, the surface $E_A$ has spherical minima $E_A/j \varepsilon=-1$ when $\chi<1$ and $\Sigma<1$ (independently of the $\Lambda-$value).

\item[II-A)]  $\vert \varphi\vert =\vert\beta\vert=\frac{\pi}{2}$
  ($E_A=-\frac{\Sigma}{2}$). The extrema do not depend on the values
  of $\Sigma$ and $\chi$. The Hessian matrix is
 \begin{equation}
  \left(
  \begin{array}{c|c}
    0  & -1 \\
    \hline
    -1 & \Sigma - \chi
  \end{array}
  \right)
  \end{equation} 
with eigenvalues:
$    \frac{1}{2} \left(\Sigma -\chi \pm \sqrt{(\Sigma-\chi)^2+4}\right).$
It turns out that both eigenvalues are always of different sign. Therefore, this solution will correspond to a saddle point.

\item[III-A)] $\beta=0$, $\cos\varphi=\frac{1}{\chi}$
  ($E_A/(j\varepsilon)=-\frac{\chi ^2+1}{2 \chi }$). Valid for
  $\chi>1$.
  The Hessian matrix is in this case
 \begin{equation}
  \left(
  \begin{array}{c|c}
    \frac{\chi^2 -1}{\chi}  & 0 \\
    \hline
    0 & \chi -  \Sigma
  \end{array}
  \right)
  \end{equation} 
with obvious eigenvalues: $\frac{\chi^2-1}{\chi}$ and  $\chi -\Sigma$.
%
%
Therefore, for
  \begin{itemize}
  \item $\chi>\Sigma$: it generates a minimum.
  \item $\chi<\Sigma$: it generates a saddle point.
  \end{itemize}
The effective order parameters are
\begin{equation}
\frac{\left\langle J^{+}\right\rangle_A }{j}=\frac{\left\langle J^{-}%
\right\rangle_A }{j}=\frac{\sqrt{\chi^2-1}}{\chi},\qquad
\frac{\left\langle A_{1}^{+}\right\rangle_A }{j}=\frac{\left\langle A_{-1}^{+}\right\rangle_A }{j}
=\frac{\left\langle A_{0}^{+}\right\rangle_A }{j}=0
\end{equation}
Assuming a small parameter $x$ such that $\chi=1+x$, the critical exponent can be shown to be $\epsilon=1/2$
\begin{equation}
\frac{\left\langle J^{+}\right\rangle_A }{j}=
\frac{\left\langle J^{-}\right\rangle_A }{j}\sim \sqrt{2x},
\end{equation}
this points towards the existence of a second order QPT.

This solution is linked to a non-zero value of the variational
parameter $\varphi$ associated to the Hartree-Fock transformation
(Eq. (\ref{lasD})), because of that we will call this solution the
Hartree-Fock (HF) deformed solution. Consequently, the surface $E_A$
has a  HF deformed minimum $E_A/j \varepsilon= - \frac{\chi^2 + 1}{2
  \chi}$ when $\chi>1$ and $\chi > \Sigma$ (independently of the
$\Lambda-$value). 

\item[IV-A)] $\varphi=0$, $\cos\beta=\frac{1}{\Sigma}$ ($E_A/(j\varepsilon)=-\frac{\Sigma ^2+1}{2 \Sigma }$)  for $\Sigma>1$.
  The Hessian matrix is in this case
 \begin{equation}
  \left(
  \begin{array}{c|c}
    \frac{\Sigma- \chi}{\Sigma^2}  & 0 \\
    \hline
    0 & \frac{\Sigma^2-1}{\Sigma}
  \end{array}
  \right)
  \end{equation} 
with obvious eigenvalues: $\frac{\Sigma- \chi}{\Sigma^2}$ and  $\frac{\Sigma^2-1}{\Sigma}$.  

Therefore, for:
\begin{itemize}
\item $\chi<\Sigma$: it generates a minimum.
\item $\chi>\Sigma$: it generates a saddle point.
\end{itemize}

Assuming a small parameter $x$ such that $\Sigma=1+x$, the critical exponent can be shown to be $\epsilon=1/2$
\begin{equation}
\frac{\left\langle J^{+}\right\rangle_A }{j}=\frac{\left\langle J^{-}%
\right\rangle_A }{j}=0,\qquad \frac{\left\langle A_{1}^{+}\right\rangle_A }%
{j}=\frac{\left\langle A_{-1}^{+}\right\rangle_A }{j}=\frac{\sqrt{\Sigma^{2}-1}}{\sqrt{2}\Sigma}\sim
\sqrt{x}.%
\end{equation}
Again, this points towards the existence of a second order QPT.

This solution corresponds to a non-zero value of the variational
parameter $\beta$ linked to the Bogoliubov transformation
(Eq. (\ref{lasv})), because of that we will call this solution the BCS
deformed solution. Consequently, the surface $E_A$ has a  BCS deformed
minimum $E_A/j \varepsilon=- \frac{\Sigma^2+1}{2\Sigma}$ when $\Sigma>1$ and $\chi < \Sigma$ (independently of the $\Lambda-$value). 

\item[V-A)] $\ \cos\beta\cos\varphi=\frac{1}{\chi}$  ($E_A/(j\varepsilon)=-\frac{\chi^2+1}{2 \chi}$) for the particular case $\chi=\Sigma$. 

Solutions III) and IV) are particular cases of this solution for
$\chi=\Sigma$. The solution corresponds to a minimum, with one of the
eigenvalues of the Hessian matrix being positive, while the other
zero, therefore this solution corresponds to a kind of closed
valley. The degeneracy of solutions III-A), IV-A), and V-A) is an
indicator of the presence of a first order QPT. 
\end{enumerate}

\subsection{Energy surface B}
This energy surface is obtained with the selection of phases as stated
in the second row of Table \ref{phases}. It can be
written as, 
\begin{equation}
E_{B}=-\varepsilon j\cos\varphi\cos\beta-2hj^{2}\sin^{2}\beta\sin^{2}%
\varphi-Vj^{2}\sin^{2}\varphi\cos^{2}\beta.
\label{eq_energ2}
\end{equation}
Again, in order to present the results, it is convenient to rescale the energy functional with the size $j$ of the system. Then, the second energy functional reads, 
\begin{equation}
\frac{E_{B}}{j\varepsilon}=-\cos\varphi\cos\beta-\Lambda\sin^{2}\beta\sin
^{2}\varphi-\frac{\chi}{2}\sin^{2}\varphi\cos^{2}\beta.
\label{eq_energ2b}
\end{equation}
(Note that the factor $-1$ in the denominator of (\ref{eq_par}) is not taken into account because a large value of $j$ is assumed). In this case, the effective  order parameters are
\begin{equation}
\frac{\left\langle J^{+}\right\rangle_B }{j}=\frac{\left\langle J^{-}%
\right\rangle_B }{j}=\sin\varphi\cos\beta,
\label{orpar_e2a}
\end{equation}
\begin{equation}
\frac{\left\langle A_{1}^{+}\right\rangle_B}{j}=\frac{\left\langle
A_{-1}^{+}\right\rangle_B}{j}=0, \qquad
\frac{\left\langle A_{0}^{+}\right\rangle_B}{j}=\sin\beta\sin\varphi.
\label{orpar_e2b}
\end{equation}

To study the extrema of (\ref{eq_energ2b}), first we imposse the
derivatives of the energy surface to be equal to zero 
\begin{eqnarray}
\frac{\partial E_{B}}{j\varepsilon \partial\beta}&=&\sin\beta\left(  \cos\varphi-(2\Lambda - \chi)
\cos\beta\sin^{2}\varphi \right)=0, \nonumber \\
\frac{\partial E_{B}}{j\varepsilon \partial\varphi}&=&\sin\varphi\left(  \cos\beta - \cos \varphi
-(2\Lambda - \chi) \sin^{2}\beta\cos\varphi \right)=0.
\label{extremaB}
\end{eqnarray}
Later, to determine the nature of the extrema, i.e., minima, maxima or saddle points, we calculate the Hessian matrix (vertical and horizontal lines are included for clarity),
\begin{equation}
\left(
  \begin{array}{c|c}
     \label{hessianB} 
{\cal H}_{\varphi,\varphi} & {\cal H}_{\varphi,\beta} \\
\hline
{\cal H}_{\beta,\varphi} & {\cal H}_{\beta,\beta}
\end{array}
\right) =
\left(
\begin{array}{c|c}
\cos (\beta ) \cos (\varphi ) - \chi \cos(2 \varphi) + & - \sin (\beta ) \sin (\varphi ) +  \\
(\chi-2 \Lambda) \sin^2 (\beta) \cos(2\varphi) &  \frac{1}{2} (\chi- 2\Lambda) \sin (2\beta ) \sin (2 \varphi) \\
\hline
- \sin (\beta ) \sin (\varphi ) +  & \cos (\beta ) \cos (\varphi ) +  \\
\frac{1}{2} (\chi- 2\Lambda) \sin (2\beta ) \sin (2 \varphi) & (\chi-2 \Lambda) \cos (2 \beta) \sin ^2(\varphi)
\end{array}
\right)
\end{equation}
The solution of the equations (\ref{extremaB}) leads to different scenarios. These are,

%
\begin{enumerate}
\item [I-B)] $\beta=0$, $\varphi=0$ ($E_B/(j\varepsilon)=-1)$, regardless the values of $\chi$ and $\Lambda$.
  In this case, the Hessian matrix is diagonal
\begin{equation}
  \left(
  \begin{array}{c|c}
   1 - \chi  & 0 \\
    \hline
    0 & 1
  \end{array}
  \right)
  \end{equation} 
with obvious eigenvalues $1-\chi$ and $1$, therefore:
\begin{itemize}
\item $\chi<1$: it generates a minimum.
\item $\chi>1$: it generates a saddle point.
\end{itemize}
Both order parameters are equal to 0. Consequently, the surface $E_B$ has spherical minima $E_B/j \varepsilon=-1$ independently of the $\chi-$, $\Sigma-$ and $\Lambda-$values.

\item [II-B)] $\beta=0$, $\cos\varphi=\frac{1}{\chi}$ ($E_2/(j\varepsilon)=-\frac{\chi ^2+1}{2 \chi }$)  for $\chi>1$.
The Hessian matrix is
\begin{equation}
  \left(
  \begin{array}{c|c}
   \frac{\chi^2-1}{\chi}  & 0 \\
    \hline
    0 & \chi + 2 \Lambda \left(\frac{1}{\chi^2} - 1 \right)
  \end{array}
  \right)
  \end{equation} 
with obvious eigenvalues $\frac{\chi^2-1}{\chi}$ and $\chi + 2 \Lambda \left(\frac{1}{\chi^2} - 1 \right)$. The first eigenvalue is always positive (remember that $\chi>1$), and consequently, 
\begin{itemize}
\item $\Lambda>\frac{1}{2}\frac{\chi^3}{\chi^2-1}$: generates a saddle point.
\item $\Lambda<\frac{1}{2}\frac{\chi^3}{\chi^2-1}$: generates a minimum.
\end{itemize}
The order parameters will be
\begin{equation}
\frac{\left\langle J^{+}\right\rangle_B }{j}=\frac{\left\langle J^{-}%
\right\rangle_B }{j}=\frac{\sqrt{\chi^2-1}}{\chi},
\end{equation}
\begin{equation}
\frac{\left\langle A_{1}^{+}\right\rangle_B}{j}=\frac{\left\langle
A_{-1}^{+}\right\rangle_B}{j}=0,\qquad
\frac{\left\langle A_{0}^{+}\right\rangle_B}{j}=0.
\end{equation}
Again, assuming a small parameter $x$ such that $\chi=1+x$, the critical exponent can be shown to be $\epsilon=1/2$. This points to the existence of a second order QPT.
\begin{equation}
\frac{\left\langle J^{+}\right\rangle_B }{j}=
\frac{\left\langle J^{-}\right\rangle_B }{j}\sim \sqrt{2x}.
\end{equation}
This solution is linked to a non-zero value of the variational
parameter $\varphi$ associated to the Hartree-Fock transformation
(Eq. (\ref{lasD})). Consequently, the surface $E_B$ has a HF deformed
minimum $E_B/j \varepsilon= - \frac{\chi^2 + 1}{2  \chi}$ when $\chi>1$ and $\Lambda < \frac{1}{2}\frac{\chi^3}{\chi^2-1}$ (independently of the $\Sigma-$value).

\item [III-B)]  $\vert \varphi\vert =\vert\beta\vert=\frac{\pi}{2}$ ($E_B/(j\varepsilon)=-\Lambda$).

The Hessian matrix is
\begin{equation}
  \left(
  \begin{array}{c|c}
   2 \Lambda  & -1 \\
    \hline
    -1 & 2 \Lambda - \chi
  \end{array}
  \right)
  \end{equation} 
with eigenvalues $\frac{1}{2} \left(4 \Lambda - \chi \pm \sqrt{\chi ^2+4} \right)$.
Therefore:
\begin{itemize}
\item $\Lambda>\frac{1}{4}(\chi+\sqrt{4+\chi^2})$: it generates a minimum.
\item
  $\frac{1}{4}(\chi+\sqrt{4+\chi^2})>\Lambda>\frac{1}{4}(\chi-\sqrt{4+\chi^2})$:
  it generates a saddle point.
\item $\Lambda<\frac{1}{4}(\chi-\sqrt{4+\chi^2})<0$: it generates a maximum.
\end{itemize}

The order parameters will be
\begin{equation}
\frac{\left\langle J^{+}\right\rangle_B}{j}=\frac{\left\langle J^{-}%
\right\rangle_B}{j}=0,
\end{equation}
\begin{equation}
\frac{\left\langle A_{1}^{+}\right\rangle_B}{j}=\frac{\left\langle
A_{-1}^{+}\right\rangle_B}{j}=0,\qquad
\frac{\left\langle A_{0}^{+}\right\rangle_B}{j}=1.
\end{equation}

\begin{itemize}
\item Solutions III-B) and II-B)   become degenerated for
  $\Lambda=\frac{1+\chi^2}{2\chi}.$ 
\item Solutions III-B) and I-B) become degenerated for $\Lambda=1$.
\item Solutions III-B) and IV-A) ({\it First energy surface})  become
  degenerated for $\Lambda=\frac{1+\Sigma^2}{2\Sigma}$. 
\end{itemize}
Therefore, the existence of this solution points towards the presence
of a first order QPT. This solution is linked to non-zero values of
both variational parameters ($\varphi$, $\beta$), one associated to
the Hartree-Fock transformation (Eq. (\ref{lasD})) and the other to
the Bogoliubov transformation (Eq. (\ref{lasv})), therefore this is a
combined HF-BCS deformed solution. Consequently, the surface $E_B$ has
a deformed HF-BCS minimum $E_B/j \varepsilon= - \Lambda$ when $\Lambda
> \frac{1}{4}(\chi+\sqrt{4+\chi^2})$ (independently of the
$\Sigma-$value). 

\item [IV-B)] The last solutions of Eqs. (\ref{extremaB}) imply
  \begin{eqnarray}
    \label{sol-IV-1}
    \cos\varphi&=&\cos\beta\sin^{2}\varphi\left(  2\Lambda-\chi\right), \\
    \cos\beta&=&\cos\varphi\left(1 +  (2\Lambda - \chi) \sin^{2}\beta \right),
                 \label{sol-IV-2}
  \end{eqnarray}
  with an energy
  \begin{equation}
    E_B/(j\varepsilon)=\frac{1-2\sqrt{2\Lambda}\sqrt{2\Lambda -\chi}}{2(2\Lambda-\chi)}.
  \end{equation}
It can be proved that the solutions always correspond to a saddle point.
\end{enumerate}

Once analyzed both surfaces, the phase diagram of the model is
obtained in the next section taking into account the competition
between both surfaces which give rise to diferent regions, some of
them including coexistence of the three phases: spherical (S), HF
deformed (HF) and BCS deformed (BCS), besides combined the combined
HF-BCS deformed solution.  

\section{Phase diagram}
\label{sec-phase-dia}
Based on the analysis of the previous section, we can derive a phase
diagram with five different phases:
\begin{itemize}
\item Symmetric or spherical solution, $(\varphi=0, \beta=0)$. It
  corresponds to solutions I-A) and I-B) (letters A or B  indicate the
  energy surface). 
\item HF deformed solution, $(\cos\varphi=\frac{1}{\chi}$ and
  $\beta=0)$. It corresponds to solutions  III-A) or  II-B). 
\item BCS deformed solutions,  ($\varphi=0$ and
  $\cos\beta=\frac{1}{\Sigma}$). It corresponds to solution IV-A). 
\item Combined HF-BCS deformed solution,
  $(\varphi=\frac{\pi}{2},\beta=\frac{\pi}{2})$. It corresponds to
  solution III-B). 
\item Closed valley ($\cos\varphi \cos \beta= 1/\chi$). It corresponds
  to solution V-A). 
\end{itemize}

\begin{figure}[hbt]
  \centering
\includegraphics[width=.5\linewidth]{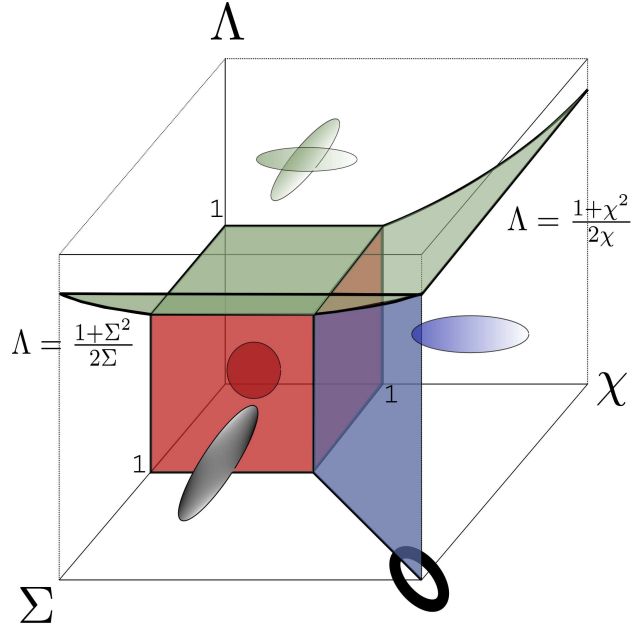}
\caption{Phase diagram of the extended Agassi Hamiltonian
  (\ref{eq_h_agassi-si}). Red vertical planes correspond to second
  order QPT surfaces. The green surface ($\Lambda=1$ for $\chi<1$ and
  $\Sigma<1$, $\Lambda=\frac{1+\chi^2}{2\chi}$ for $\chi>\Sigma$ and
  $\Lambda=\frac{1+\Sigma^2}{2\Sigma}$ for $\chi<\Sigma$) and the blue
  vertical one ($\chi=\Sigma$ and $\Lambda <
  \frac{1+\Sigma^2}{2\Sigma}$) correspond to first order critical
  surfaces. Red sphere, blue oval, black oval, black thick oval, and
  crossed green ovals stand for the symmetric solution, the HF
  deformed solution, the BCS deformed solution, the V-A) solution, and
  HF-BCS deformed solution, respectively.} 
\label{fig-phase-dia}
\end{figure}

In Fig.~\ref{fig-phase-dia} we depict the phase diagram of the
model. The phase diagram is built considering that two energy surfaces
(A and B) coexist and compite but only one (except when they are
degenerate) gives the absolute minimum.  Concerning $E_A$, one has to
take into account that their minima only depend on $\chi$ and $\Sigma$
but they do not depend on $\Lambda$, while regarding $E_B$, their
minima only depends on $\chi$ and $\Lambda$ but they do not depend on
$\Sigma$. The region with $\chi<1$, $\Sigma<1$, and $\Lambda<1$
corresponds to the symmetric phase and it is represented in the
diagram with a red sphere. The red vertical surface $\chi=1$ with
$\Lambda\le 1$ and $\Sigma \le 1$ is a second order QPT. This can be
shown easily looking at the energy values for $\chi<1$ which is $E=-1$
and for $\chi>1$ which is $E= -\frac{1+\chi^2}{2 \chi}$. These
expressions imply a discontinuity in the second order derivative of
the energy at $\chi=1$. The other red vertical surface $\Sigma=1$ with
$\Lambda\le 1$ and  $\chi \le 1$, is also a second order QPT as can be
shown by an equivalent argument. The area at the right bottom
corresponds to the HF deformed shape and it is depicted with a blue
prolate shape oriented in the direction of the $\chi$ axis. The area
at the left bottom corresponds to the BCS deformed shape and it is
represented with a black prolate shape oriented in the direction of
the $\Sigma$ axis. The vertical blue plane with $\chi=\Sigma$
corresponds to a first order QPT since at this particular surface
solutions III-A) and IV-A) are degenerated. Besides, solution V-A)
exists, it corresponds to a closed valley in the $\beta-\varphi$ plane
and it is represented with a black thick oval in the figure. Finally,
the region above the green surface corresponds to the solution
$(\varphi=\frac{\pi}{2}, \beta=\frac{\pi}{2})$, i.e., to the combined
HF-BCS
deformed solution and it is represented in the diagram with two
crossed green ovals. The green surface corresponds to a first order QPT
since all the energies below the green surface are
$\Lambda$ independent while above $E=-\Lambda$. 

In Fig.~\ref{fig-phase-dia} we represent the
deepest minimum of the lowest energy surface. However, because of the
presence of two competing energy surfaces, $E_A$ and $E_B$, there are
areas where different phases coexist. These regions have not been
depicted in Fig.~\ref{fig-phase-dia} because of the complexity that
would generate in the phase the diagram. For $E_A$ the spherical, the
HF deformed and the BCS deformed solutions cannot coexist. However,
for $E_B$ different phase can coexist: i) the spherical and the
combined HF-BCS deformed shape, and ii) the combined HF-BCS and the HF
deformed. Moreover, the competition between $E_A$ and $E_B$ produces
the coexistence of up to five different minima: 

\begin{itemize}
\item Spherical $(\varphi=0,\beta=0)$, HF-BCS deformed minimum
$(\varphi=\pi/2, \beta=\pi/2)$ and BCS deformed one $(\varphi=0,\beta
\arccos(1/\Sigma))$.
\item HF-BCS deformed minimum $(\varphi=\pi/2, \beta=\pi/2)$, HF
deformed minimum in $(\varphi=\arccos(1/\chi),\beta=0)$ and BCS
deformed one in $(\varphi=0,\beta=\arccos(1/\Sigma))$.
\item Closed valley minimum (for $\Sigma=\chi$) and the combined
HF-BCS deformed minimum $(\varphi=\pi/2, \beta=\pi/2)$.
\item HF-BCS deformed minimum $(\varphi=\pi/2, \beta=\pi/2)$, HF
deformed minimum in $(\varphi=\arccos(1/\chi),\beta=0)$, BCS deformed
minimum and the closed valley minimum along the line $\Lambda= \frac{1
+ \chi^2}{2 \chi}$ with $\chi=\Sigma$. All the minima are degenerated.
\item Spherical $(\varphi=0,\beta=0)$, HF-BCS deformed minimum
$(\varphi=\pi/2, \beta=\pi/2)$, BCS deformed one $(\varphi=0,\beta
\arccos(1/\Sigma))$, HF deformed minimum in
$(\varphi=\arccos(1/\chi),\beta=0)$, and the closed valley minimum at
the point $\chi=\Sigma=\Lambda=1$. All the minima are degenerated.
\end{itemize}

\begin{figure}[hbt]
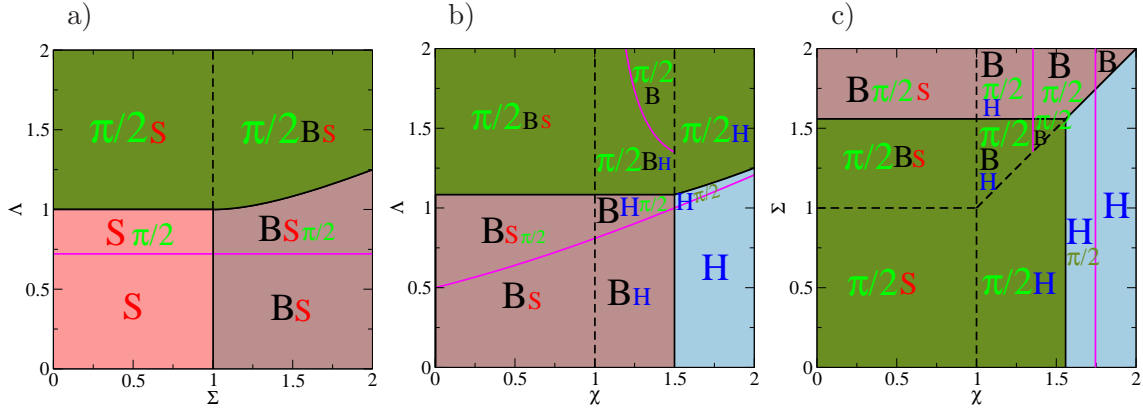

  \begin{tabular}{ccc}
a)\hspace{3cm} & b)\hspace{3cm} & c)\hspace{3cm} \\
    \includegraphics[width=.3\linewidth]{cut-chi0.75.eps}&
    \includegraphics[width=.3\linewidth]{cut-sigma1.5.eps}&
    \includegraphics[width=.3\linewidth]{cut-lambda1.1.eps}
  \end{tabular}
    \caption{Phase diagram for selected planes a) $\chi=0.75$, b)
      $\Sigma=1.5$, and c) $\Lambda=1.1$. Background color represents
      the shape of the deepest minimum, i.e., full light red for the symmetric
      phase, full olive green for combined HF-BCS deformed solution, full
      brown for BCS deformed phase, and full light blue for HF deformed
      phase. Letters correspond to the existing phases, i.e., red
      ``S'' for symmetric phase, green ``$\pi/2$'' for combined HF-BCS
      deformed solution, black ``B'' for BCS deformed phase, and blue
      ``H'' for HF deformed phase. Larger and left most (or upper)
      letters correspond to deepest minima. Black full lines stand for
      QPTs, magenta full ones for spinodal-antispinodal lines, and
      black dashed ones for indicating the change in the ordering of
      high lying minima.} 
\label{fig-phase-cuts}
\end{figure}

In order to have a more clear image of the phase diagram, in
Fig.~\ref{fig-phase-cuts} we present 2-dimensional plots in which one
of the control parameters has been kept constant. In panel (a) we
depict the plane $\chi=0.75$. Since $\chi<1$ the spherical (red S)
minimum I-B) always exists. For $\Sigma> 1$ the BCS deformed minimum
(black B) also exists, IV-A). For
$\Lambda>\frac{1}{4}(\chi+\sqrt{4+\chi^2})$, which is the spinodal
line (full magenta line) of $E_B$, the HF-BCS combined minimum (green
$\pi/2$) appears, III-B). This results in different areas of
coexistence as shown in the figure. The largest most left letter gives
the deepest minimum, the second most left letter indicates the second
deepest minimum, an so on. For instance, in the region $\Sigma>1$ and
$\Lambda> \frac{1+\Sigma^2}{2 \Sigma}$ three phases coexist: the
HF-BCS deformed is the lowest, then the BCS deformed and higher in
energy the spherical phase. In the region just below, the same three
phases coexist but now the lowest is the BCS deformed, then the
spherical and higher in energy is the combined HF-BCS deformed
solution.

In panel (b) of Fig.~\ref{fig-phase-cuts} we present the case
$\Sigma=1.5$. For $\chi<1$ the spherical phase, I-B), always is
present. For $\chi<\Sigma$ the BCS deformed minimum, IV-A),
exists. For $\Lambda>\frac{1}{4}(\chi+\sqrt{4+\chi^2})$, which is the
spinodal line (lowest full magenta line) of $E_B$, the HF-BCS combined
minimum (green $\pi/2$) appears, III-B). For $\chi>\Sigma$ the HF
solution, III-A), always exists. In addition, in the region
$1<\chi<\Sigma$ and $\Lambda< \frac{1}{2}\frac{\chi^3}{\chi^2-1}$,
which is the anti-spinodal line (upper full magenta line), the HF
deformed solution, II-B), also exists. Again, different coexistence
regions appear as marked in the figure following the same criteria as
in a).

In panel (c) of Fig.~\ref{fig-phase-cuts} we analyze the vertical
plane $\Lambda=1.1$. We consider this particular case because it
corresponds to a range of parameters in which the most complex
situation exists. For $\chi<1$ the spherical minimum, I-B), always
exists. For $\chi<\Sigma$ and $\Sigma>1$ the BCS deformed solution,
IV-A), exists. For $\chi < \frac{4 \Lambda^2-1}{2 \Lambda}$, right
most magenta line, (which is obtained from the spinodal line
$\Lambda=\frac{1}{4}(\chi+\sqrt{4+\chi^2})$) the combined HF-BCS
deformed exists, III-B). For $\chi>\Sigma$ and $\chi>1$ the HF
deformed solution exists, III-A). In addition, there exists the HF
deformed solution, II-B), for $\chi>1$ and $\Sigma>\chi$ and
fulfilling the condition related to the anti-spinodal line,
$\Lambda<\frac{1}{2}\frac{\chi^3}{\chi^2-1}$. Different coexistence
regions appear as marked in the figure following the same criteria as
in a).

In summary, the phase diagram presents four regions where the shapes
are S, HF deformed, BCS deformed, and combined HF-BCS deformed,
respectively. However, in each region several minima exist, coexisting
up to three phases in certain regions. In addition, there is a line,
$\chi = \Sigma$ with $\Lambda=\frac{1+\chi^2}{2\chi}$, in which four
phases, HF, BCS, HF-BCS and the closed valley solutions, are
degenerated, plus a single point, $\chi=\Sigma=\Lambda=1$, in which
the five solutions (same as before plus the spherical) are
degenerated. Such a rich phase diagram do not appear even in the case
of more complex systems, such as the proton-neutron interacting boson
model \cite{Aria04}, the two-fluid Lipkin model \cite{Garc16} or for
Hamiltonians with up two three-body interactions \cite{Fort11}.
\begin{figure}[hbt] \centering
\begin{tabular}{cc}
\includegraphics[width=.7\linewidth]{energ-order-par-15.eps} &
\includegraphics[width=.3\linewidth]{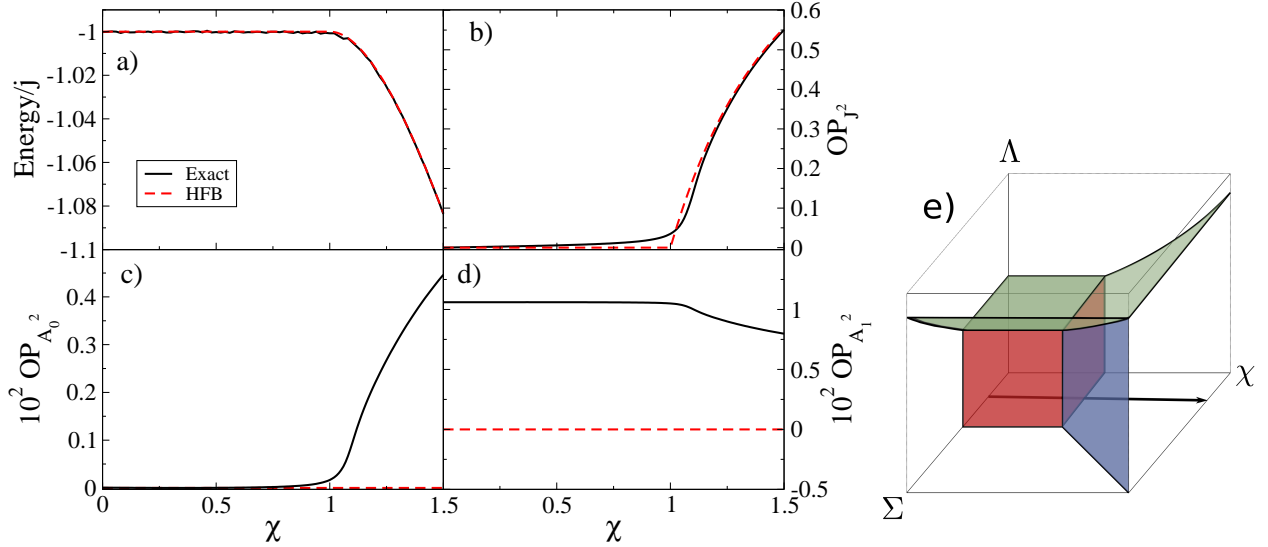}
\end{tabular}
\caption{Comparison of the ground-state energy per fermion pair and
order parameter values for HFB and exact results for a system with
$j=100$ and Hamiltonian parameters $\Sigma=0.5$, $\Lambda=0$
($\varepsilon=1$) as a function of $\chi$. Black full lines correspond
to exact results and red dashed ones to HFB ones. Panel a) corresponds
to the ground state energy, panel b) to $OP_{J^2}$, panel c) to
$OP_{A_0^2}$, panel d) to $OP_{A_1^2}$ order parameters, and panel e)
to the schematic representation of the trajectory in the parameter
space. Please, note that the scale in panels c) and d) are multiplied
by 100.}
\label{energ-order-par-15}
\end{figure}

\section{Comparison between exact and Hartree-Fock-Bogoliubov results}
\label{sec-comparison}
In this section we present several cases where we calculate by an
exact diagonalization of the Hamiltonian the values of the  ground-state energy and of different effective order parameters, and compare them with the HFB results.

We define the effective order parameters in terms of the expectation values of the following operators for the ground state, although, in general, they can be used with excited states,
\begin{eqnarray}
\label{ef_j} OP_{{J}^2}&=&\frac{\langle (J^+)^2 \rangle + \langle (J^-)^2\rangle}{2 j^2},\\
\label{ef_a0}OP_{{A}^2_0}&=&\frac{\langle A_0^+A_0\rangle}{j^2},\\
\label{ef_a2}OP_{{A}^2_1}&=&\frac{\langle A_1^+A_1\rangle +\langle A_{-1}^+A_{-1}\rangle}{2 j^2}.
\end{eqnarray}
Note that these quantities differ from
Eqs.~(\ref{orpar_e1a},\ref{orpar_e1b}) or
Eqs.~(\ref{orpar_e2a},\ref{orpar_e2b}) since the expectation values of
single operators $J^+$, $J^-$, $A_0$, $A_1$, and $A_{-1}$ vanish owing
to parity conservation in the exact solution. Therefore,
Eqs.~(\ref{ef_j}, \ref{ef_a0}, \ref{ef_a2}) should be compared with  the square of either Eqs.~(\ref{orpar_e1a},\ref{orpar_e1b}) or Eqs.~(\ref{orpar_e2a},\ref{orpar_e2b}).

All the diagonalizations presented in this section are performed for systems with $j=100$ ($100$ pairs of fermions), $\varepsilon=1$ and only positive parity states are considered. This number of fermion pairs is large enough to guarantee a good agreement between the HFB mean field values and the exact ones.

\begin{figure}[hbt]
  \centering
\begin{tabular}{cc}
\includegraphics[width=.7\linewidth]{energ-order-par-10b.eps} &
\includegraphics[width=.3\linewidth]{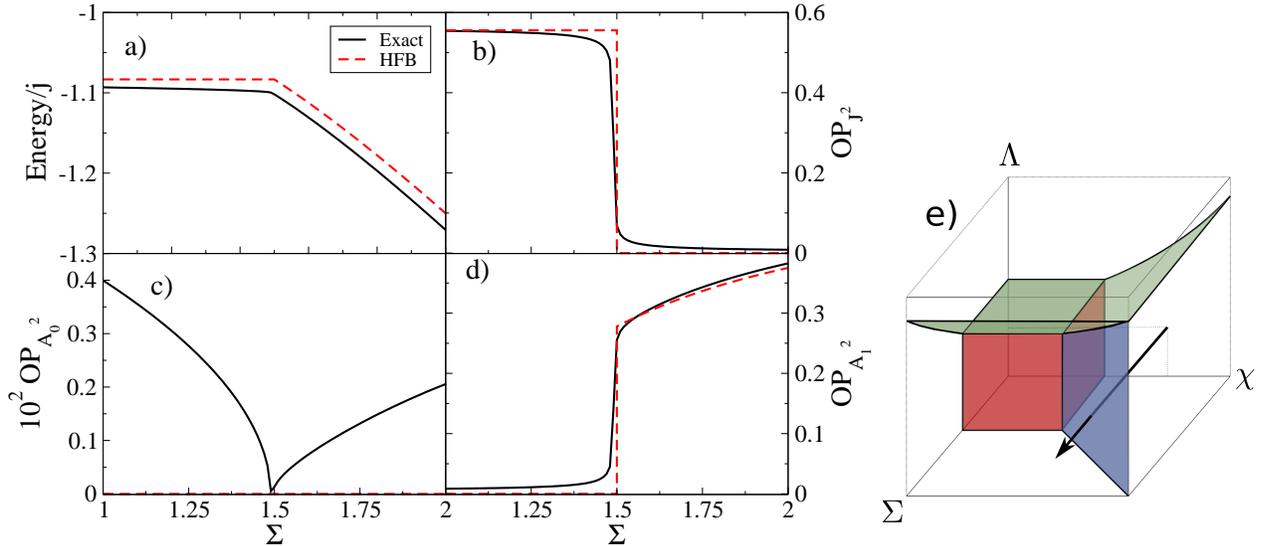}
\end{tabular}
\caption{Same caption as in Fig.~\ref{energ-order-par-15} but for $\chi=1.5$, $\Lambda=0.5$, as a function of $\Sigma$. Please, note that the scale in panel c) is multiplied by 100.}
\label{energ-order-par-10}
\end{figure}

First, we consider a trajectory that goes through one of the red
vertical surfaces and, therefore, should correspond to cross a second
order QPT. In particular, in Fig.~\ref{energ-order-par-15} we depict
such a situation, for which we fix the parameters $\Sigma=0.5$ and
$\Lambda=0$, allowing to vary the value of $\chi$ between $0$ and
$1.5$ (see panel (e) for the schematic trajectory). All along this
trajectory $\beta=0$, while $\varphi=0$ for $\chi<1$ and
$\varphi=\arccos\frac{1}{\chi}$ for $\chi>1$. This means that we
explore the transition between spherical and  HF deformed shapes. In
panel (a), the ground state energy is shown as a function of $\chi$,
suggesting the presence of a second order QPT around $\chi=1$. The HFB
ground state energy is $-1$ for $\chi<1$ (spherical phase) and
$E/(j\varepsilon)=-\frac{\chi ^2+1}{2 \chi }$ for $\chi>1$ (HF
deformed phase). As stated before, this involves a discontinuity in
the second derivative of the energy with respect to $\chi$. In panel
(b), the effective order parameter $OP_{{J}^2}$ (\ref{ef_j}) is
depicted, and a good agreement between HFB and exact results is
obtained. The HFB value is $OP_{{J}^2}=0$ for $\chi<1$, while
$OP_{{J}^2}=1-\frac{1}{\chi^2}$ for $\chi>1$. On the other hand, in
spite of the good  agreement found with the exact calculations, we can
observe a small discrepancy around the critical point $\chi=1$. This
is due to the finite size of the system. In order to describe properly
that region, it is needed to take into account corrections to the size
of the system that would improve the results obtained with the HFB
approach. In panel (c) and (d) the HFB value of $OP_{A_0^2}$ (\ref{ef_a0}) and  $OP_{A_1^2}$ (\ref{ef_a2}) are depicted, respectively, and their analytical values are zero all the way. Note that the vertical scale of these two panels has been multiplied by a factor $100$, which can lead the reader to the impression that the agreement is poor, which is not the case because the absolute difference between the exact and analytical results is up to order $10^{-2}$. Finally, it is worth noting that this QPT only involves minima of the first energy surface, $E_A$.

In Fig.~\ref{energ-order-par-10} we move through the horizontal line
$\chi=1.5$, $\Lambda=0.5$, as a function of $\Sigma$, crossing the
$\chi=\Sigma$ vertical plane (see panel (e)), going from the minimum
$(\varphi=\arccos(1/\chi)$, $\beta=0)$ (HF deformed) to $(\varphi=0$,
$\beta=\arccos(1/\Sigma))$ (BCS deformed). As we can see, this QPT
transition is of first order, because the the HFB energy changes from
a constant value $E/(j\varepsilon)=-\frac{1.5^2+1}{2\cdot 1.5^2}$ for
$\Sigma<1.5$ to $E/(j\varepsilon)=-\frac{\Sigma^2+1}{2\Sigma^2}$ for
$\Sigma>1.5$, with a sudden jump from one minimum to the other,
although there is no coexistence. In the case of the HFB value of $OP_{{J}^2}$ (\ref{ef_j})(panel (b)), the expectation value changes from $OP_{{J}^2}=1-\frac{1}{1.5^2}$ for $\Sigma<1.5$ to zero for $\Sigma>1.5$, presenting a discontinuity in the order parameter $OP_{{J}^2}$ at $\Sigma=1.5$. On the other hand, in panel (c) one can see how the HFB value for $OP_{A_0^2}$ (\ref{ef_a0}) is strictly zero for all $\Sigma-$values, while very small values are obtained in the exact calculation (note that the vertical scale is multiplied by a factor 100). Finally, in panel (d) the mean field value for $OP_{A_1^2}$ (\ref{ef_a2}) jumps from zero to $OP_{A_1^2}=1-\frac{1}{\Sigma^2}$ at $\Sigma=1.5$. The exact calculation follows, except for finite number of particle corrections the same behaviour. Note that this QPT only involves minima of the first energy surface, $E_A$.

\begin{figure}[hbt]
  \centering
\begin{tabular}{cc}
\includegraphics[width=.7\linewidth]{energ-order-par-12b.eps} &
\includegraphics[width=.3\linewidth]{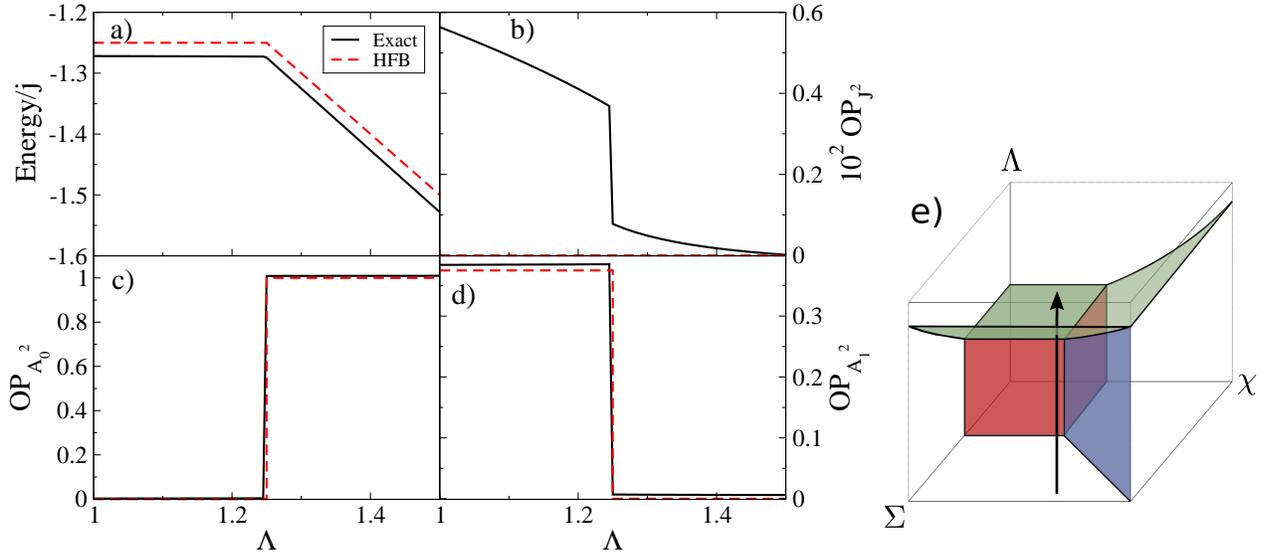}
\end{tabular}
\caption{Same caption as in Fig.~\ref{energ-order-par-15} but for $\chi=1.5$, $\Sigma=2$, as a function of $\Lambda$. Please, note that the scale in panel b) is multiplied by 100.}
\label{energ-order-par-12}
\end{figure}

Finally, in Fig.~\ref{energ-order-par-12} we move through a vertical
line with $\chi=1.5$, $\Sigma=2$, as a function of $\Lambda$, crossing
the surface $\Lambda=\frac{1+\Sigma^2}{2\Sigma}$ (see panel (e)), then
passing from  $(\varphi=0, \beta=\arccos(1/\Sigma))$ (deformed BCS) to
$(\varphi=\pi/2,\beta=\pi/2)$ (combined deformed HF-BCS). The QPT
appears at $\Lambda=\frac{1+2^2}{4}=1.25$. As can be seen in the
energy per fermion pair shown in panel (a), the QPT is, once more, of
first order and the HFB energy passes from
$E/(j\varepsilon)=-\frac{2^2+1}{2\cdot 2^2}$ to
$E/(j\varepsilon)=-\Lambda$ in $\Lambda=1.5$. In panel (b), the mean
field HFB value for $OP_{{J}^2}$ (\ref{ef_j}) is zero all along the
path, while the exact calculation gives a very small value (note that
the vertical scale is multiplied by a factor $100$). In panel (c) the
HBF mean field value for $OP_{A_0^2}$ (\ref{ef_a0}) jumps suddenly at
$\Lambda=1.25$ from zero to  $1$. The same behaviour is obtained in
the exact calculation. In panel (d), the HFB mean field value for
$OP_{A_1^2}$ (\ref{ef_a2}) jumps suddenly from
$OP_{A_1^2}=\frac{2^2-1}{2\cdot 2^2}=0.375$ to zero at $\Lambda=1.25$,
and, once more, the exact calculation gives a consistent result. Note
that this QPT involves the change from a minimum of the energy surface
$E_A$, to another one of the energy surface $E_B$. 

All the above results confirm the structure of the phase diagram shown in Fig.\ \ref{fig-phase-dia} as well as the character, first or second order, of the QPTs.

\section{Summary and conclusions}
\label{sec-conclusions}
We have presented an extended version of the Agassi model which includes the extra $A_0^\dag A_0$ contribution in the Hamiltonian and, therefore, it has four free parameters, $\varepsilon$, $g$, $V$, and $h$, although we have always considered a non vanishing value for $\varepsilon$, hence the number of effective free parameters is three: $V=\frac{\varepsilon\chi}{2j-1}$, $g=\frac{\varepsilon\Sigma}{2j-1}$, and $h=\frac{\varepsilon\Lambda}{2j-1}$. We have performed a HFB mean field approach and we have got the corresponding energy surfaces. It is worth to be noted that two different energy surfaces appear, each one depending only on two of the control parameters. The existence of two different energy surfaces is due to the freedom in the election of the phase in the Bogoliubov transformation.

We have analyzed the equilibrium value of the order parameters,
$\varphi$ and $\beta$, for minima, maxima and saddle points and we
have settled the phase diagram of the model. In the phase diagram four
regions can be distinguished: symmetric, HF deformed, BCS deformed,
and HF-BCS deformed phases. Moreover, there is a special situation in
which the HF and the BCS deformed minima are correlated, plane
$\chi=\Sigma$. We have called this situation closed valley minimum. In
the four regions different phases can coexist, in fact, there are
regions with up to three coexisting phases. In addition, there is a
line in which four phases coexist and are degenerated plus a single
point $\chi=\Sigma=\Lambda=1$ in which the five phases are degenerated
(spherical, HF deformed, BCS deformed, combined HF-BCS deformed, and
close valley deformed minimum). The ground state is completely
determined by the lowest energy minimum of the lower energy surface in
each region. The existence of other minima does not affect the ground
state properties, but it is expected to have a strong influence in the
presence of excited-state quantum phase transitions \cite{Garc17}.

Finally, we have compared the exact results with the HFB mean field values for different observables. In all the cases, good agreement has been obtained validating the mean field results.

The phase diagram of the present extended Agassi model shows a rich variety of phases. Phase coexistence is present in extended areas of the parameter space. The model could be an important tool for benchmarking novel many-body approximations.

\section{Acknowledgment}
This work has been supported by the Spanish Ministerio de
Econom\'{\i}a y Competitividad and the European regional development
fund (FEDER) under Projects No. FIS2014-53448-C2-1/2-P and
FIS2015-63770-P, and by Consejer\'{\i}a de Econom\'{\i}a,
Innovaci\'on, Ciencia y Empleo de la Junta de Andaluc\'{\i}a (Spain)
under Groups FQM-160 and FQM-370.


\begin{thebibliography}{99}
\bibitem{Agas68} D.\ Agassi, Validity of the bcs and rpa
  approximations in the pairing-plus-monopole solvable model, Nucl.\
  Phys.\ A {\bf 116}, 49 (1968).

\bibitem{Lipk65} H.J.~Lipkin, N.~Meshkov, and A.J.~Glick, Validity of
  many-body approximation methods for a solvable model. (I). Exact
  solutions and perturbation theory, Nucl.\ Phys.\ {\bf 62}, 188 (1965).

\bibitem{Davi86} E.D.\ Davis and W.D.\ Heiss, Random-phase
  approximation and broken symmetry, J.\ Phys.\ G: Nucl.\ Phys.\ {\bf
    12}, 805 (1986).

\bibitem{Iach06} F.~Iachello, {\it Lie Algebras and Applications}
  (Springer-Verlag, Berlin Heidelberg, Germany, 2006).

\bibitem{Jayn63} E.T.~Jaynes and F.W.~Cummings, Comparison of
  quantum and semiclassical radiation theories with application to the
  beam maser, Proc.\ IEEE {\bf 51}, 89 (1963).

\bibitem{Dick54} R.H.~Dicke, Coherence in Spontaneous Radiation
  Processes, Phys.\ Rev.\ {\bf 93}, 99 (1954).

\bibitem{Hoga61} J.\ Hogaasen-Feldman, A study of some approximations
  of the pairing force, Nucl.\ Phys.\ {\bf 28}, 258 (1961).

\bibitem{Elli58} J.P.\ Elliott, Collective motion in the nuclear shell
  model. I. Classification schemes for states of mixed configurations,
  Proc.\ Roy.\ Soc.\ (London) A {\bf 245}, 128 (1958).

\bibitem{Iach87} F.~Iachello and A.~Arima, {\it The Interacting Boson Model}
(Cambridge University Press, Cambridge, UK, 1987).

\bibitem{Cejn09} P.\ Cejnar and J.\ Jolie, Quantum phase transitions
  in the interacting boson model, Prog.\ Part.\ Nucl.\ Phys.\ {\bf 62}, 210 (2009).

\bibitem{Cejn10} P.\ Cejnar, J.\ Jolie, and R.F.\ Casten, Quantum
  phase transitions in the shapes of atomic nuclei, Rev.\ Mod.\ Phys.\
  {\bf 82}, 2155 (2010).

\bibitem{Schu69} K.\ Bleuler, A.\ Friederich, and D.\ Schutte,
  Validity of the hartree-bogoliubov-theory in an exactly solvable
  model, Nucl.\ Phys.\ A {\bf 126}, 628 (1969).


\bibitem{Gert83} Y. Gerstenmaier, Radius of convergence of the
  hole-line expansion and of the Rayleigh-Schr\"odinger perturbation
  series in a solvable model, Nucl.\ Phys.\ A {\bf 394}, 16 (1983).

\bibitem{Herm17} M. R. Hermes, J. Dukelsky and  G. E. Scuseria,
  Combining symmetry collective states with coupled cluster theory:
  Lessons from the Agassi model Hamiltonian, Phys.\ Rev.\ C {\bf 95}, 064306
  (2017).


\bibitem{Hirs97} J.G.\ Hirsch, P.O.\ Hess, and O. Civitarese, Single-
  and double-beta decay Fermi transitions in an exactly solvable
  model, Phys.\ Rev.\ C {\bf 56}, 199 (1997).
  
\bibitem{Aria04} J.M.\ Arias, J.E.\ Garc\'{\i}a-Ramos, and J.\
  Dukelsky. ``Phase Diagram of the Proton-Neutron Interacting Boson
  Model'', Phys.\ Rev.\ Lett.\ {\bf 93},  212501 (2004).
  
\bibitem{Garc16} J.E.\ Garc\'{\i}a-Ramos, P.\ P\'erez-Fern\'andez,
  J.M.\ Arias, and E.\ Freire, ``Phase diagram of the two-fluid Lipkin model: A
  ``butterfly'' catastrophe'', Phys.\ Rev.\ C {\bf 93}, 034336 (2016). 
  
\bibitem{Fort11} L.\ Fortunato, C.E.\ Alonso, J.M.\ Arias, J.E.\
  Garc\'{\i}a-Ramos, and A.\ Vitturi, ``Phase diagram 
  for a cubic-Q interacting boson model Hamiltonian: Signs of
  triaxiality'', Phys.\ Rev.\ C {\bf 84}, 014326 (2011). 

\bibitem{Garc17} J.E.\ Garc\'{\i}a-Ramos, P.\ P\'erez-Fern\'andez, and 
  J.M.\ Arias, ``Excited-state quantum phase transitions in a
  two-fluid Lipkin model'', Phys.\ Rev.\ C {\bf 95}, 054326 (2017). 
   
\end{thebibliography}
\end{document}